\documentclass[12pt,preprint]{aastex}
\usepackage{natbib}
\shortauthors{Thommes} \shorttitle{Interactions between Type I and
II planets}

\title{A safety net for fast migrators:  Interactions between gap-opening and sub-gap-opening bodies in a protoplanetary disk}
\author{Edward W. Thommes}
\affil{Canadian Institute for Theoretical Astronomy, University of
Toronto, 60 St. George Street, Toronto, Ontario M5S 3H8}
\email{thommes@cita.utoronto.ca}

\begin{abstract}
Young planets interact with their parent gas disks through tidal
torques.  An imbalance between inner and outer torques causes
bodies of mass $\ga 0.1$ Earth masses ($M_\oplus$) to lose angular
momentum and migrate inward rapidly relative to the disk; this is
known as ``Type I'' migration. However, protoplanets that grow to
gas giant mass, O($10^2)\,M_\oplus$ , open a gap in the disk and
are subsequently constrained to migrate more slowly, locked into
the disk's viscous evolution in what is called ``Type II"
migration. In a young planetary system, both Type I and Type II
bodies likely coexist; if so, differential migration ought to
result in close encounters when the former originate on orbits
exterior to the latter. We investigate the resulting dynamics,
using two different numerical approaches: an N-body code with
dissipative forces added to simulate the effect of the gas disk,
and a hybrid code which combines an N-body component with a
1-dimensional viscous disk model, treating planet-disk
interactions in a more self-consistent manner. In both cases, we
find that sub-gap-opening bodies have a high likelihood of being
resonantly captured when they encounter a gap-opening body. A
giant planet thus tends to act as a barrier in a protoplanetary
disk, collecting smaller protoplanets outside of its orbit.  Such
behavior has two important implications for giant planet
formation: First, for captured protoplanets it mitigates the
problem of the migration timescale becoming shorter than the
growth timescale when $M_{\rm proto}\ga 1 M_\oplus$. Secondly, it
suggests one path to forming systems with multiple giant planets:
Once the first has formed, it traps (or indeed accretes) the
future solid core of the second in an exterior mean-motion
resonance, and so on.  The most critical step in giant planet
formation may thus be the formation of the very first one.
\end{abstract}
\begin{document}
\section{Introduction}
Observational evidence strongly suggests that the formation of
giant planets is a ubiquitous process.
To date, well over a hundred extrasolar planets have been
discovered by radial velocity surveys; in at least eleven cases,
multiple planets orbit a single star\footnote{IAU Working Group on
Extrasolar planets,
http://www.ciw.edu/boss/IAU/div3/wgesp/planets.shtml; California
and Carnegie Planet Search, http://www.exoplanets.org}.
At least three of the multi-planet systems have pairs of planets
locked in mean-motion resonances (MMRs), with a 1:2 (GJ 876b,c; HD
82943b,c) or 1:3 (55 Cnc b,c) period ratio.

Migration due to planet-disk interaction is the leading candidate
for the mechanism which produces such dynamical configurations, as
well as producing the large fraction of ``hot Jupiters",
exoplanets on orbits significantly smaller than their likely
formation distance from their parent star. Indeed, the latter
effect was first predicted, and investigated in detail, well
before it was observed \citep{1980ApJ...241..425G,
1986ApJ...309..846L}.
More recent work has expanded the focus from planet-disk
interaction to ``planet-planet-disk" dynamics, that is, the
behavior of multiple interacting planets embedded in a disk.
Again, theoretical work anticipated observation:  Using
hydrodynamic simulations, \cite{2000MNRAS.313L..47K} and
\cite{2000ApJ...540.1091B} demonstrated convergent planet
migration, thus raising the possibility of resonant capture, just
prior to the discovery of the first extrasolar MMR, the 1:2
resonant pair of GJ876b and c.
Subsequent work
\citep*{2001MNRAS.320L..55M,2002ApJ...567..596L,2002ApJ...565..608M,2002MNRAS.333L..26N,2004A&A...414..735K}
looked in more detail on the dynamics of capture and evolution of
gas giant planets in low-order MMRs.

The above work focused on the interaction of pairs of {\it
gap-opening} planets, that is, planets which have grown large
enough to  lock themselves into the viscous evolution of their
parent disk.  For typical protoplanetary disk parameters, this
implies planets of gas giant (Jovian) size, O($10^2$) Earth masses
($M_\oplus$), rather than ice giant (Neptunian) size,
O($10^1)\,M_\oplus$ planets.   For initially well-separated
planets, convergent migration then requires the removal of the
intervening gas, thus bringing the planets to share a common gap.
It is possible, however, for giant planet MMRs to be established
earlier.  Prior to opening a gap, disk tidal torques can cause the
orbital decay of an embedded body on a timescale much shorter than
the disk's viscous time.  This mode of migration is commonly
referred to as Type I, while a gap-opening planet co-evolving with
its disk is said to undergo Type II migration
\cite{1997Icar..126..261W}.
\cite{1996LPI....27..479H} suggested that the potentially large
differences between Type I and Type II timescales could lead to
bodies of the former class being delivered to, and then resonantly
trapped by, bodies of the latter class.  The concentration of a
significant mass of protoplanets in exterior MMRs of a
pre-existing giant planet could, they argued, promote the
formation of subsequent giant planets.  In this work, we undertake
the first part of a numerical investigation of this scenario.  We
make use of two different approaches.  In the first and simpler
one, we take a standard N-body code and add to it forces to mimic
the effects of planet-disk interaction, guided by analytic work as
well as by numerical simulations.  The second approach is to take
this same N-body code and add to it a simple disk model, which
viscously co-evolves with embedded planets and interacts with
them.  The disk model is one-dimensional; that is, it describes an
azimuthally-averaged thin disk, with disk-planet torques obtained
from analytic formulae rather than from a full calculation of the
launching of density waves by the planet. This approach is
intermediate between the above ``N-body with disk forces" and an
actual 2- (or 3-) dimensional hydrodynamic simulation.  With both
approaches, we find that for typical protoplanetary disk
parameters, single Type I bodies have a high likelihood of being
resonantly captured by Type II bodies.  We further demonstrate
that significant numbers of protoplanets can occupy adjacent
resonances, or indeed share a single resonance, of a giant planet.
This paper is organized as follows:  In \S \ref{background}, we
give a brief summary of the theory of planet-disk interactions. In
\S \ref{omni}, we present simulations performed using the simpler
of our two approaches, while in \S \ref{vdisk} we repeat these
simulations using our hybrid ``N-body+viscous disk" code. We
discuss our results in \S \ref{discussion}, and summarize them in
\S \ref{summary}.

\section{Planet-disk torques, migration and eccentricity
evolution} \label{background}

Here, we provide a brief summary of the theory of planet-disk
interaction, focusing on the results we will make use of in this
paper. A body orbiting in a gas disk launches density waves at
Lindblad resonances, exchanging energy and angular momentum with
the disk; the disk thus modifies the body's orbit. The net effect
on the body can be obtained by summing over all resonances
(\citealp*{1980ApJ...241..425G}; see also e.g.
\citealp*{2003ApJ...585.1024G}). One begins by expanding the
gravitational potential of the perturbing body in a double Fourier
series.  The $l,m$ component has a pattern speed
\begin{equation}
\Omega_{l,m} = \Omega_p + (l-m) \kappa_p/m
\end{equation}
where $\Omega_p$ and $\kappa_p$ are the azimuthal and epicyclic
frequencies, respectively, of the planet. There is a Lindblad
resonance between this potential component and disk material at a
radius $r$ in the disk where
\begin{equation}
\Omega(r) \pm \frac{\kappa(r)}{m} = \Omega_{l,m}.
\end{equation}
Also, a corotation resonance occurs where
\begin{equation}
\Omega(r) = \Omega_{l,m}
\end{equation}
 The radial pressure
gradient in the disk causes both the epicyclic and the azimuthal
frequency to differ slightly from their Keplerian values:
\begin{equation}
\Omega^2(r) = \frac{G M_*}{r^3} + \frac{1}{r \rho}
\frac{\partial}{\partial r}(\rho c_s^2) \label{pressure supported
omega}
\end{equation}
and
\begin{equation}
\kappa^2(r) = \frac{1}{r^3} \frac{\partial}{\partial r}\left ( r^4
\Omega^2 \right ) \label{pressure supported kappa}.
\end{equation}

In the case of a planet on a circular orbit, all components with
$l \ne m$ are zero; $\Omega_{m,m} = \Omega_p$, so all components
have the pattern speed of the planet.  Furthermore, there is no
torque contribution at all from corotation resonances.  We will
focus on this simpler situation for the moment. In summing the
effect of all Lindblad resonances, one can define a torque
density, or torque per unit disk radius. Following
\citet*{1997Icar..126..261W}, the torque density experienced by
the disk due to a planet of mass $M=\mu M_*$ on a circular orbit
about a primary of mass $M_*$ at radius $r_p$, is
\begin{equation}
\left [ \frac{dT}{dr}(r) \right ]_{LR} = {\rm sgn}(r-r_p) \frac{2
\mu^2 \Sigma(r) r_p^4 \Omega_p^4}{r(1+4 \xi^2)\kappa^2} m^4 \psi^2
\label{torque density}
\end{equation}
where $\xi \equiv m c_s/r \kappa$, $c_s$ is the gas sound speed,
and $\Psi$ is the dimensionless satellite forcing function,
\begin{equation}
\Psi = \frac{\pi}{2} \left [ \frac{1}{m} \left | \frac{d
b^m_{1/2}}{d \beta} \right | + 2 \frac{\Omega}{\kappa} \sqrt{1 +
\xi^2} \beta^m_{1/2}(\beta)\right ],
\end{equation}
with $\beta \equiv r/r_p$ and
\begin{equation}
b^m_{1/2}(\beta) = \frac{2}{\pi} \int^{\pi}_0 \frac{\cos m \theta
d \theta}{\sqrt{1 - 2 \beta \cos \theta + \beta^2}}
\end{equation}
being the Laplace coefficient. Following
\citet*{2004ApJ...606..520M}, we account in a simple, approximate
way for the finite thickness of the disk by replacing the Laplace
coefficient with a softened approximation,
\begin{equation}
b^m_{1/2}(\beta) \approx \frac{2}{\pi \beta^{1/2}}K_0 \left ( m
\sqrt{\beta - 2 + \frac{1}{\beta} + \frac{(B H)^2}{r r_p}} \right
)
\end{equation}
with $K_0$ the modified Bessel function of the second kind, order
0, $H \approx c_s/\Omega$ the disk scale height, and $B$ a
dimensionless scaling factor.  In going from a summing of torques
at discrete resonances to a torque density, the wavenumber $m$ is
turned into a continuous function of radius:
\begin{equation}
m(r) = \left [
\frac{\kappa^2}{(\Omega-\Omega_p)^2-c_s^2/r^2}\right ]^{1/2}.
\end{equation}

\citet{1988Icar...73..330W}, \citet{1993ApJ...419..166A} and
\citet{2000MNRAS.315..823P} study planet-disk torques from an
eccentric planet. In particular,
\citet{2000MNRAS.315..823P} look at the case where eccentricity
exceeds disk aspect ratio H/r.  They
 sum over enough $l \ne m$ torques for
convergence, in order to obtain the net effect on the planet's
orbit.  However, they neglect corotation torques, as these are
zero for their model surface density profile ($\Sigma \propto
r^{-3/2}$).  They obtain approximate fits for the timescales of
angular momentum and eccentricity change:
\begin{equation}
t_m \equiv -\frac{J}{\dot{J}}= 3.5 \times 10^5 f_s^{1.75} \left
[\frac{1+\left ( \frac{e_p a_p}{1.3 H}\right )^5}{1 - \left
(\frac{e_p r_p}{1.1 H} \right )^4}\right ] \left (
\frac{H/a_p}{0.07}\right )^2 \left ( \frac{\Sigma}{588{\rm
gcm^{-2}}}\right )^{-1} \left ( \frac{M_p}{1 M_\oplus}\right
)^{-1} \left ( \frac{a_p}{1 AU}\right )^{-1/2} {\rm yrs}
\label{pap t_m}
\end{equation}
and
\begin{equation}
t_e \equiv -\frac{e}{\dot{e}}= 2.5 \times 10^3 f_s^{2.5} \left [
1+\frac{1}{4}\left (\frac{e_p}{H/a_p} \right )^3\right ] \left (
\frac{H/a_p}{0.07}\right )^2 \left ( \frac{\Sigma}{588{\rm
gcm^{-2}}}\right )^{-1} \left ( \frac{M_p}{1 M_\oplus}\right
)^{-1} \left ( \frac{a_p}{1 AU}\right )^{-1/2} {\rm yrs}
\label{pap t_e}
\end{equation}
In the limit of low planetary eccentricities, these results are
broadly consistent with the earlier work of
\citet{1997Icar..126..261W} and \citet{1993ApJ...419..166A}.
  Most noteworthy is the short orbital decay timescale; for
constant eccentricity, $a/\dot{a} = t_m/2$, so from Eq. \ref{pap
t_m}, it is of order $10^4$ years for a body of mass 10 $M_\oplus$
in a typical model disk. This would seem to present a significant
problem for the core-accretion model of giant planet formation,
since the assembly of the solid core likely takes a million years
or more
(e.g. \citealp{1987Icar...69..249L},
\citealp{2000Icar..143...15K}, \citealp*{2003Icar..161..431T},
\citealp*{2003Icar..166...46I}), with perhaps several million more
required to reach runaway gas accretion (itself a very fast
process) and acquire  a Jovian atmosphere
\citep{1996Icar..124...62P}.

If a planet does manage to grow to something approaching gas giant
mass before plunging into its parent star, the problem of rapid
inward migration becomes less severe.  Torque increases with
planet mass; for a sufficiently massive body, $\sim 10^2 M_\oplus$
for typical disk models, the torques overcome disk viscosity and
an annular gap about the orbit is opened in the gas
\citep{1980ApJ...241..425G}, provided that the Hill (Roche) radius
also exceeds the disk scale height
(\citeauthor{1986ApJ...309..846L} \citeyear{1986ApJ...309..846L},
\citeyear{1993prpl.conf..749L}). A planet opening a clean gap
prevents disk material from crossing its orbit, and thus becomes
locked into the disk's viscous evolution. In general, this entails
a sharp decrease in the rate of orbital decay.

The eccentricity evolution of a gap-opening planet is as yet not
well understood.  In the gapless case, the dominant effect on the
planet's eccentricity is damping by coorbital Lindblad resonances.
When a gap forms and empties or strongly depletes the coorbital
region, the evolution of the eccentricity comes down to a delicate
balance between the effect of non-coorbital eccentric Lindblad
resonances (driving) and corotation resonances (damping). Though
the latter are slightly stronger than the former,
\citet{2003ApJ...585.1024G} and \citet{2003ApJ...587..398O} argue
that corotation resonances in fact saturate, thus favouring net
eccentricity excitation. However,
\citet{2003ApJ...585.1024G} suggest that damping may be
reestablished once a planet's eccentricity becomes so high that
its radial excursion exceeds the gap width. In contrast, numerical
simulations tend to show only damping of eccentricities
(e.g. \citealp*{2004A&A...414..735K}) unless the planet mass
approaches that of a brown dwarf
\citep*{2001A&A...366..263P}.

\section{Numerical simulations}
\label{simulations}

For the purpose of the subsequent numerical experiments we will
assume that, the above difficulties notwithstanding, a
protostellar disk does manage to form one initial gas giant
planet, massive enough to open a gap and orbitally co-evolve with
its parent disk.  Taking this as our initial condition, we then
study the evolution of sub-gap-opening bodies originating at
larger radii than the gas giant.  To better assess the robustness
of our results, we repeat the investigation with two different
numerical approaches.

\subsection{N-body + dissipative disk forces}
\label{omni}

Simulations are performed with a code based on SyMBA (Symplectic
Massive Body Algorithm), an N-body program optimized for
near-Keplerian systems
\citep*{1998AJ....116.2067D}.  SyMBA is based on the symplectic
orbit-integration algorithm of
\citet{1991AJ....102.1528W}, with its high efficiency and good
energy-conservation properties, and adds an adaptive timestep in
order to resolve close encounters between bodies.  Using a model
gas disk, we then calculate the angular momentum and eccentricity
decay rate for each sub-gap-opening body from Eqs. \ref{pap t_m}
and \ref{pap t_e} each timestep, and apply the corresponding
additional acceleration $\vec{a} = a_{\phi} \hat{\phi} + a_r
\hat{r}$, where
\begin{equation}
a_{\phi} = -\frac{\dot{J}}{M_p r_p} = \frac{1}{t_m} \frac{J}{M_p
r_p}
\end{equation}
and
\begin{equation}
a_{r} = -2 \frac{\vec{v} \cdot \vec{r}} {r^2 t_e}.
\end{equation}
The giant planet in our simulations is treated differently,
however. For simplicity, we prevent it from migrating altogether.
This is done in a way that mimics, rather simplistically, the
effect of being locked in a disk gap: We choose a radius $r_{\rm
gap}$ for the center of the gap and a fractional gap ``width" $w$,
then apply an azimuthal acceleration $a_{\phi}'$ as follows:
\begin{equation}
a_{\phi}'= \frac{a-r_{\rm gap}}{|a-r_{\rm gap}|+w a}a_{\phi}
\end{equation}
Since $a_{\phi}$ is negative, $a_{\phi}'$ is negative at $a>r_{\rm
gap}$, positive at $a<r_{\rm gap}$, and falls off smoothly to zero
as $a \rightarrow r_{\rm gap}$.  Therefore, if our ``gap-opening"
planet is displaced from the gap midpoint, it experiences a
restoring torque towards $a=r_{\rm gap}$.

\subsubsection{Single growing protoplanet}
\label{single core omni}

\begin{figure}
\plotone{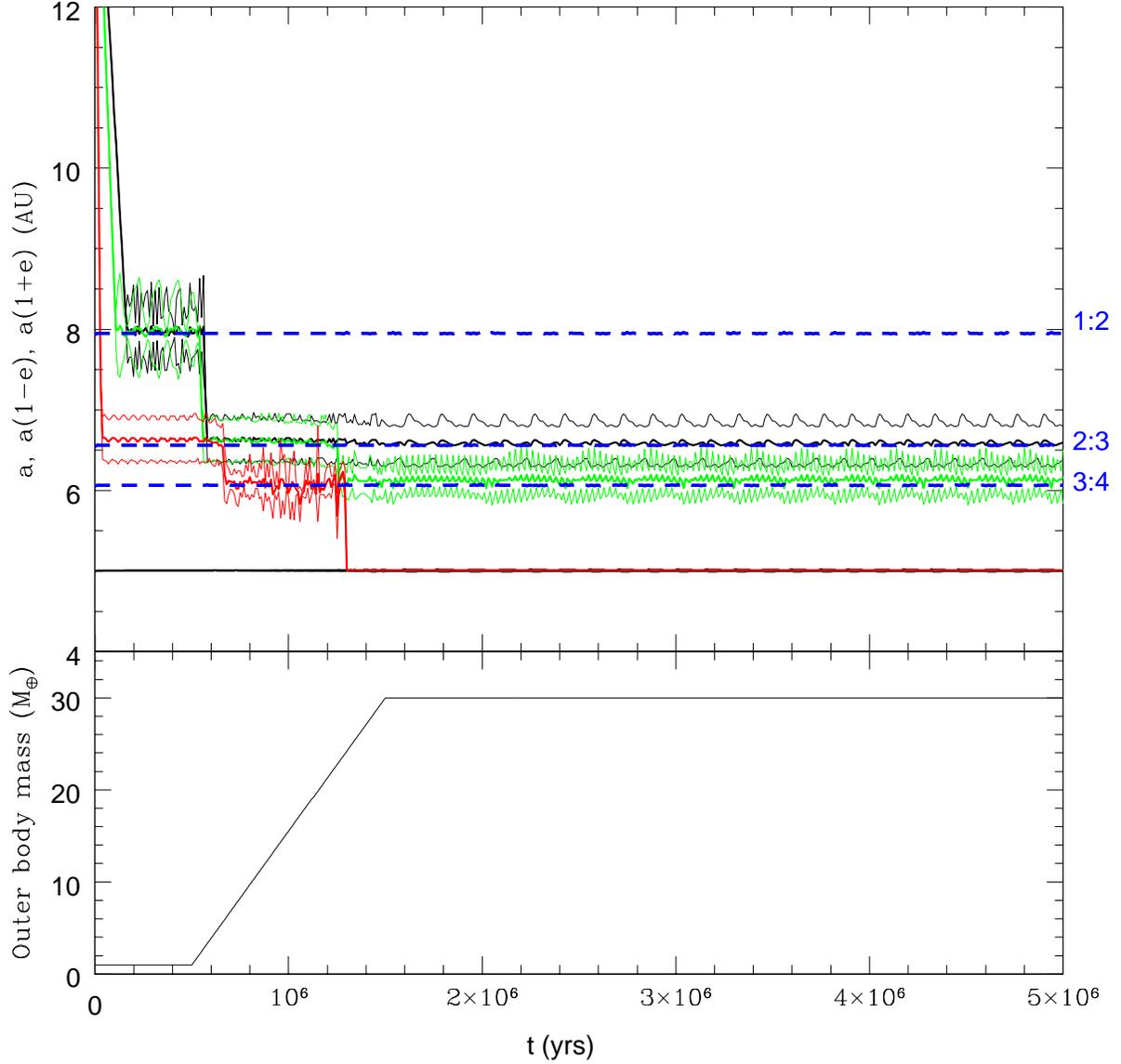}
\caption{Simulation of growing, sub-gap-opening (Type I) body in a
gas disk, using the ``N-body + dissipative disk forces'' scheme.
The simulated disk has a surface density of $\Sigma_{\rm 1 AU}
(r/{\rm AU})^{-1}\,{\rm g cm^{-2}}$.  In the top panel, the
orbital evolution is shown for three cases: $\Sigma_{\rm 1
AU}=1000\,{\rm g \, cm^{-2}}$ (black), $\Sigma_{\rm 1
AU}=1500\,{\rm g \,cm^{-2}}$ (green), and $\Sigma_{\rm 1
AU}=6000\,{\rm g \, cm^{-2}}$ (red).  In each case the body's
semimajor axis $a$, pericenter $q \equiv a(1-e)$ and apocenter $Q
\equiv a(1+e)$ are plotted.  A body of mass 300 $M_\oplus$ orbits
at 5 AU; the Type I body spends time trapped in several of its
mean-motion resonances (blue dashed lines).  The time evolution of
the Type I body's mass is the same in each of the three cases, and
is shown in the bottom panel.} \label{single_core_omni}
\end{figure}

We begin with a simple case:  A 300 $M_\oplus$ (approximately
Jupiter-mass) body starts on a circular orbit at 5 AU about a 1
$M_\odot$ star, centered in a ``disk gap" as described above.  We
place a second body of mass 1 $M_\oplus$ on a circular orbit at 15
AU.  A model disk with surface density
\begin{equation}
\Sigma_{\rm g} = 1000 \left ( \frac{r}{\rm 1 AU} \right
)^{-1}\,{\rm g\,cm^{-2}} \label{baseline Sigma_g}
\end{equation}
and scale height
\begin{equation}
H=0.03 \left ( \frac{r}{\rm 1 AU}\right )^{5/4} r \label{baseline
H}
\end{equation}
is used to calculate $dJ/dt$ and $de/dt$.  After $5 \times 10^5$
years of simulation time, the outer body's mass is increased
linearly in time to 30 $M_\oplus$ over the next million years. The
resulting orbital evolution is shown in Fig.
\ref{single_core_omni}. At its original mass of 1 $M_\oplus$, the
outer body migrates inward until it is captured in the larger
body's 1:2 exterior mean motion resonance.  Once its mass has
reached $M \approx 2.5 M_\oplus$, at $t \approx 5.5$ Myrs, it
jumps from the 1:2 to the 2:3 MMR.  It remains there through the
rest of its growth and until the end of the simulation at 5 Myrs,
showing no signs of orbital instability. We perform other
simulations with different migration rates; we achieve this by
simply scaling the disk surface density uniformly, while keeping
the disk power law index and scale height profile the same (note
that $J/\dot{J} \propto H^2 \Sigma_g^{-1}$ for small
eccentricity).  Some of these other results are also shown in Fig.
\ref{single_core_omni}.  When $\Sigma({\rm 1 AU}) = 1500\,{\rm
g\,cm^{-2}}$, the growing body undergoes a second resonance jump,
from the 2:3 to the 3:4 MMR.  When $\Sigma({\rm 1 AU}) =
6000\,{\rm gcm^{-2}}$, the body does the same, then also leaves
the 3:4 resonance, but no closer resonance is able to capture it,
and it gets all the way to the gap-opener's orbit (and in the case
pictured, physically collides with it).  Our experiment thus
suggests that, for parameters relevant to a protoplanetary
disk---$\Sigma_g \sim (10^2 - 10^3)\,{\rm g\,cm^{-2}}$, $H/r \sim
(10^{-2} - 10^{-1}) r$---single bodies of gas giant core/ice giant
mass, $\sim (10^0 - 10^1)\,M_\oplus$, have a high likelihood of
being captured upon encountering a first-order (n:n+1) mean-motion
resonance of a gas giant-sized body.

\subsubsection{Multiple protoplanets}
\label{multiple cores omni}

We now turn to the question of what happens when multiple
sub-gap-opening bodies encounter a gas giant.  Dropping several
fully-formed Neptune-mass planets into a
 disk makes for a rather questionable initial condition; as in the
 previous section, we therefore begin with small bodies whose
 masses we increase over time.  This time, we do so in a manner which, at least in a
 broad qualitative sense, conforms to how planet formation
 proceeds.  We make use of the prescription for post-runaway
 or ``oligarchic" \citep{1998Icar..131..171K} growth in the form given by
 \citet*{2003Icar..161..431T}.  This
 describes the mass accretion rate of a mass $M_{pp}$ protoplanet growing together with an
 oligarchy of similar neighbours, all embedded in a disk of much
 smaller planetesimals of characteristic mass $m_{plsml}$:
 \begin{equation}
\frac{dM_{pp}}{dt}\approx A M_{pp}^{2/3} \left ( \Sigma_s-B
M^{2/3} \right ), \label{oligarchy}
 \end{equation}
 where $\Sigma_s$ is the initial local
 surface density of the planetesimal disk, and $A$ and $B$ are
 given by
 \begin{equation}
 A=\frac{3.9 b^{2/5}C_D^{2/5} G^{1/2} M_*^{1/6} \rho_g^{2/5}}{\rho_{plsml}^{4/15} \rho_{pp}^{1/3} a_{pp}^{1/10}
 m_{plsml}^{2/15}},\,B=\frac{0.23 M_*^{1/3}}{b a_{pp}^2}
\end{equation}
$C_D$ is the drag coefficient, a dimensionless constant $\approx
1$; $b$ is the spacing in Hill radii  between adjacent
protoplanets; $M_*$ is the mass of the central star; $\rho_g
\approx \Sigma_g/H$ is the (volume) density of gas; and
$\rho_{pp}$ and $\rho_{plsml}$ are the densities of protoplanet
and planetesimals, respectively.

We put protoplanets, initially of mass 0.1 $M_\oplus$, on circular
orbits between 8 and 20 AU.  Their masses are changed over time
according to Eq. \ref{oligarchy}, with $\Sigma_s$ multiplied by a
randomly-generated factor between 0.25 and 1.75 in order to mimic
the stochastic nature of the accretion.  $\Sigma_g$ and $H$ are
given by Eqs. \ref{baseline Sigma_g} and \ref{baseline H}. We use
$b=10$, $\rho_{pp}=\rho_{plsml}=1.5\,{\rm gcm^{-3}}$,
$m_{plsml}=10^{-18}\,M_\oplus$; the latter two amount to a
planetesimal size of $\sim 0.01$ km.

\begin{figure}
\plotone{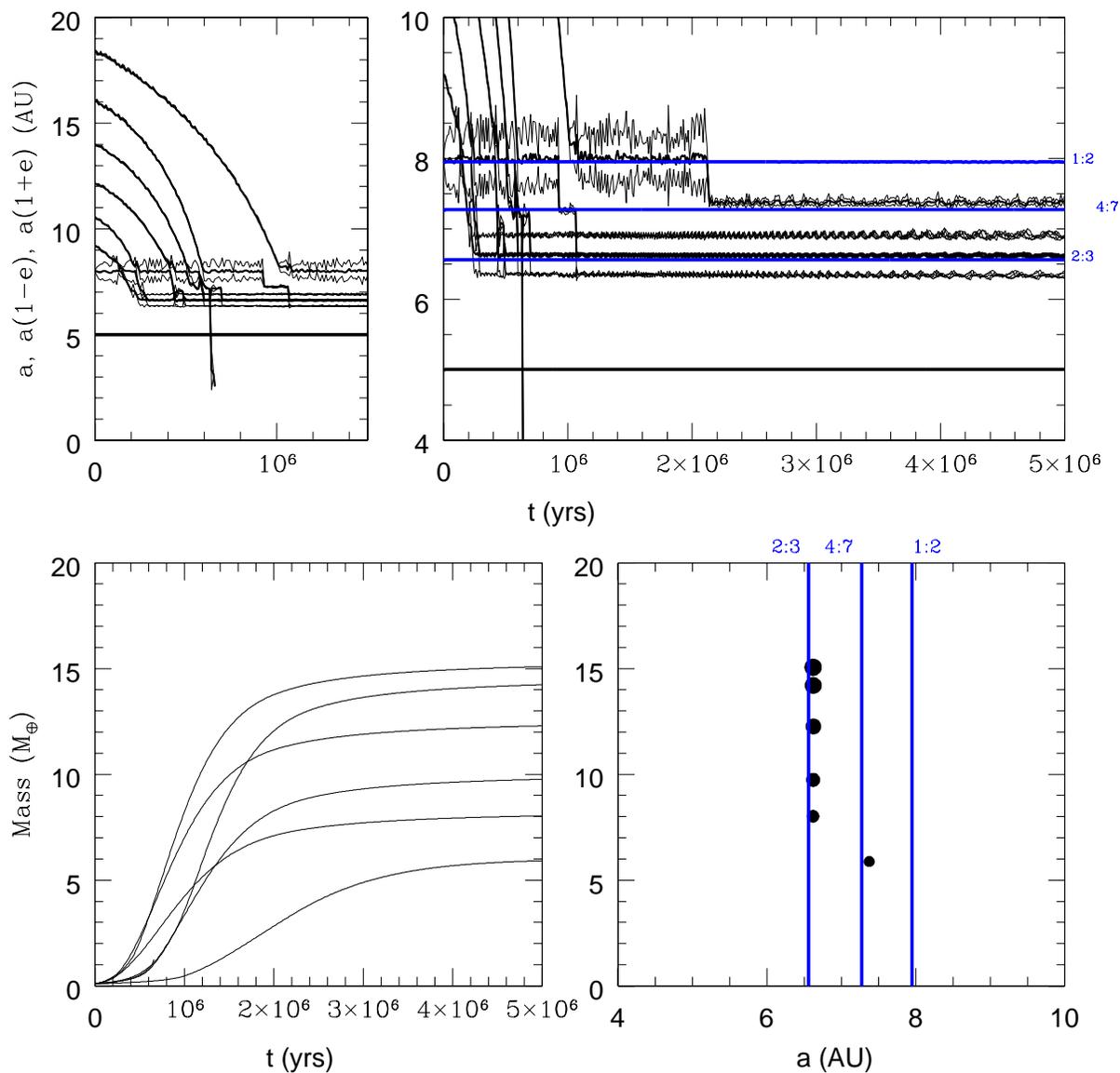}
\caption{A simulation of multiple growing protoplanets outside a
giant planet (fixed at 5 AU) using the ``N-body + dissipative disk
forces'' scheme. Top panels:  Orbital evolution of semimajor axis,
peri- and apocenters of the bodies.  The left panel shows the
initial migration, while the one on the right shows the full
simulation with a a smaller vertical scale.  Bottom left:
Evolution of the protoplanet masses by the ``oligarchy"
prescription. Bottom right: Final state showing resonance
locations.} \label{multiple_cores_omni}
\end{figure}

\begin{figure}
\plotone{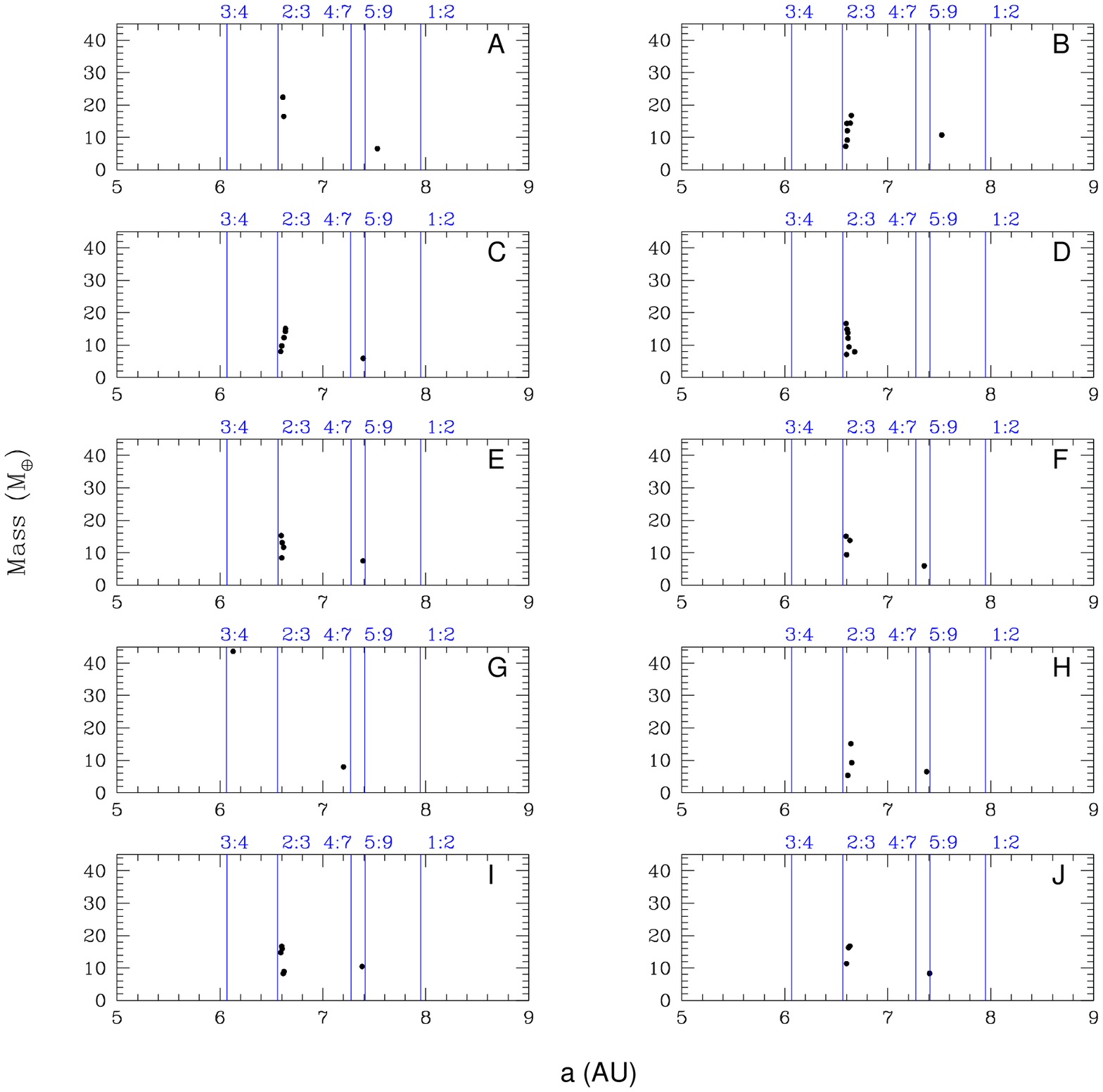}
\caption{Snapshots taken at $5 \times 10^6$ years of a set of ten
simulations (differing only in the initial orbital phases of the
bodies) of a population of growing protoplanets on orbits exterior
to a gap-opening gas giant. The locations of some of the giant's
exterior mean-motion resonance centers are shown (vertical lines).
Most of the surviving protoplanets are locked in the 2:3
resonance, typically with one additional body in the 4:7 and 5:9
resonance. In one case (G), there is a single much larger
protoplanet, formed from the merger of several smaller ones, in
the 3:4 resonance.} \label{omni_snapshots}
\end{figure}

A typical example of the orbital evolution which results is shown
in Fig. \ref{multiple_cores_omni}.  The outer bodies accelerate
their inward migration as they grow in mass, and one after another
is resonantly captured by the ``Jupiter" at 5 AU.  As they
continue to grow and the applied torque increases, some bodies
jump to a closer-in resonance, analogous to the behavior seen in
Fig. \ref{single_core_omni}.  At the end of the simulation at 5
Myrs, five bodies share the 2:3 MMR with the gas giant.  A sixth
body is further outward; examination of the different possible
resonance angles
(e.g. \citealp{1999..Murray..Dermott..book}) shows it to be in a
4:7 MMR with the gas giant, and simultaneously, in a 6:7 MMR with
the bodies which are themselves in the 3:2 resonance with the
giant ($4:7 = 2:3 \times 6:7$). Meanwhile, all the 2:3 bodies are
in a 1:1 commensurability with each other.  This seemingly
precarious configuration is attainable here because the tendency
of the
 bodies to gravitationally scatter each other as they approach to such close range, is kept in check by
 the strong damping of eccentricities which is simultaneously
 applied.  Other outcomes are shown in Fig.
\ref{omni_snapshots}.  The 2:3 resonance is strongly favoured as
the final location of the growing protoplanets, with the smallest
bodies tending to end up further out, in a 5:9 (5:6) or 4:7 (6:7)
MMR with the gas giant (the 2:3 MMR bodies), respectively.

\subsection{Hybrid N-body + viscous disk}
\label{vdisk}

We now repeat our experiments with a different numerical approach.
The code used here is called VDisk-SyMBA; as the name suggests, it
also uses SyMBA for evolving planetary bodies, but the disk is
treated in a more sophisticated way:  We co-evolve with the
planets a one-dimensional disk model, which not only changes in
time due to its own viscosity, but also exchanges angular momentum
with the N-body part, modifying its own density profile as well as
the orbits of the planets.  A similar computational approach,
though without the N-body component, was utilized by
\citet{1986ApJ...309..846L} to follow the evolution of single
protoplanets in a disk. Since radius is its only dimension, the
disk is of necessity vertically and azimuthally averaged. However,
both vertical and azimuthal structure are {\it implicitly}
included in the model. For the former, we simply assign a scale
height; for the latter, planet-disk interaction---a process which
after all works by raising azimuthally asymmetric perturbations in
the disk--- is handled by way of the torque density formulation of
Eq. \ref{torque density} applied to the azimuthally averaged disk
surface density.  This approach to simulating planet-disk
interactions is thus intermediate in complexity between our first
method and a full hydrodynamic simulation.  The fidelity of the
planet-disk interactions is only as good as the torque density
used, but because the disk is one-dimensional, this code has a
significant speed advantage over two- or three-dimensional
hydrodynamic codes.  In practice we find that millions of years
can typically be simulated in one day of real time on a
desktop-class machine, thus simulations spanning the entire
lifetime of a protoplanetary gas disk, constrained observationally
to be $\sim 10^7$ years or less
\citep*{2001ApJ...553L.153H} are within easy reach.

The gas disk is divided into radial bins of arbitrary size at the
start of the simulation.  The most stringent upper limit on bin
size comes from the necessity of properly resolving planet-disk
torques; accurately tracking the disk's viscous self-evolution is
much more forgiving in that respect. We generally find it
necessary to scale the bin size to be some fraction of the local
disk scale height.

The governing equation for the surface density of a Keplerian
viscous disk is
\begin{equation}
\frac{\partial \Sigma}{\partial t} =
\frac{1}{r}\frac{\partial}{\partial r} \left [ 3 r^{1/2}
\frac{\partial}{\partial r} (\nu \Sigma r^{1/2}) -
\frac{r^{1/2}}{\pi \sqrt{G M_*}} \frac{\partial T}{\partial r}
\right ]
\end{equation}
where $\partial T/\partial r$ is the torque density in the disk
due to the planet (Eq. \ref{torque density}). For the viscous
self-evolution of the disk, we solve this equation each timestep,
with the second term set to zero, using an explicit upwind
differencing scheme. The planet-disk torques are computed and
applied to both planet and disk in a separate step. An important
part of this is to compute $\Omega$ in each bin; although we
assume perfectly Keplerian gas velocities for the disk's viscous
evolution, the torque asymmetry felt by sub-gap-opening bodies is,
as mentioned in \S \ref{background}, primarily due to the
sub-Keplerian gas orbital speed shifting resonance locations. We
calculate $\Omega$ in each bin at each timestep using the
finite-difference version of Eq. \ref{pressure supported omega}.
We then make the approximation that $\kappa=\Omega$.

The planet-disk torques are obtained using the expression for
torque density, Eq. \ref{torque density}.  We are thus computing
the torques for a circular orbit; when $e \ne 0$, we  use a
simplified version of Eq. \ref{pap t_e} (without the
eccentricity-dependent part in square brackets) to separately
change the eccentricity. For gas surface density $\Sigma_g$, we
average among the value at the semimajor axis, pericentre and
apocentre of the orbit.  This is a somewhat ad-hoc prescription
for the eccentricity evolution, but it does reproduce the
analytically-obtained results for a non-gap-opening body, while
for a gap-opening planet, damping of eccentricities does not
commence until the planet's radial excursion causes it to overrun
its gap.  Thus we have implemented a compromise of sorts between
numerical results on the one hand
(e.g. \citealp*{2001A&A...366..263P, 2004A&A...414..735K}) and the
analytic work of
\citet{2003ApJ...585.1024G} and \citet{2003ApJ...587..398O} on the
other.  It should also be pointed out that implicit in our
approach is the assumption that angular momentum is deposited in
the disk right at the location of a resonance.  This is a good
assumption as long as the torquing planet has $r_H$ greater than
the scale height $H$ of the disk, in this case waves launched at
resonances will indeed be nonlinear and shock immediately, thus
depositing their angular momentum right
away\citep{1993prpl.conf..749L}. However, for bodies having $r_H <
H$, a more accurate treatment would include modelling the nonlocal
deposition of angular momentum in the disk
\citep*{1996ApJ...460..832T,2002ApJ...572..566R}.  The effect of
our simplification is that we underestimate the gap-opening
threshold mass for bodies with $r_H < H$.

\subsubsection{Single protoplanet}

\begin{figure}
\epsscale{0.9} \plotone{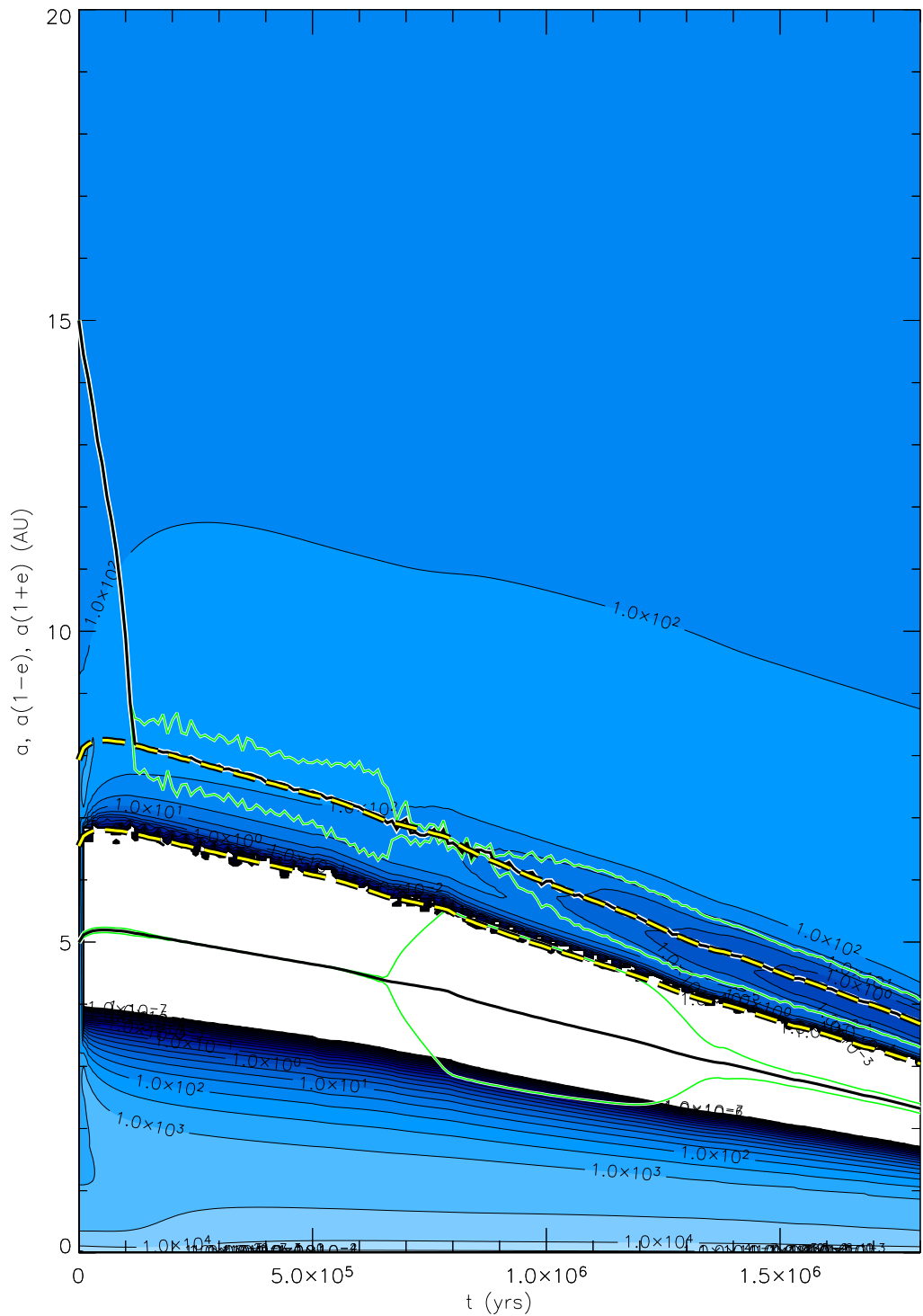}
\caption{Orbital evolution of an inner gap-opening (Type II) 300
$M_\oplus$ planet and an outer sub-gap-opening (Type I), initially
1 $M_\oplus$ body.  For each body, the semimajor axis $a$ (black),
pericenter $q\equiv a(1-e)$ and apocenter $Q \equiv a(a+e)$ (both
green) are plotted.  The locations of the gap-opening body's 1:2
and 2:3 mean-motion resonances (MMRs) are also shown (upper and
lower yellow dashed line, respectively).  The 1 $M_\oplus$ body
grows linearly to 30 $M_\oplus$ between $5 \times 10^5$ yrs and
$1.5 \times 10^6$ yrs; it remains trapped in the 1:2 MMR until the
end of the simulation.  The gap-opening body temporarily becomes
eccentric, its radial excursion filling the gap.  The evolution of
the one-dimensional disk model's surface density is overlaid as a
contour plot.  The initial surface density is $\Sigma \approx 1000
(r/{\rm 1\,AU})^{-1}\,{\rm g cm^{-2}}$.} \label{idl_paper}
\end{figure}

We generate an initial gas disk of similar surface density and
scale height profile as in \S \ref{single core omni}.  We adopt a
standard $\alpha$-parameterization for the viscosity of the disk,
$\nu = \alpha c_s H$, with $\alpha=10^{-4}$.  A 300 $M_\oplus$
body is, again, started on a circular orbit at 5 AU, while a 1
$M_\oplus$ body is started on a circular orbit at 15 AU. As
before, after $5 \times 10^5$ years, its mass is linearly
increased to 30 $M_\oplus$ over the next $10^6$ years.  The
orbital evolution of both bodies, together with the disk, is shown
in Fig. \ref{idl_paper}

The most obvious difference in the evolution, with respect to Fig.
\ref{single_core_omni}, is that the inner planet also migrates. It
immediately opens a gap in the disk, initially moving outward
slightly as the nearby disk material rearranges itself about the
freshly-opened gap. Thereafter, the planet moves inward along with
the accretion flow of the disk, going from 5 AU to 2 AU in just
under 2 Myrs.  The outer planet migrates much faster, at a similar
rate as for the $\Sigma_g({\rm 1 AU})=1000\,{\rm g\,cm^{-2}}$ case
in \S \ref{single core omni}, and is captured into the 1:2 MMR of
the inner planet after a bit more than $10^5$ years. Unlike the
cases shown in Fig. \ref{single_core_omni}, it remains in the 1:2
resonance as it grows.  The strength with which the outer body is
pushed toward the inner body is somewhat reduced by the inward
migration of the inner body, as well as by the time evolution of
the gas disk ($\Sigma_g$ decreases as the disk spreads). Also, as
can be clearly seen in Fig. \ref{idl_paper}, as the outer body
becomes more massive (from $\sim 8 \times 10^5$ years onward) it
begins to open a gap of its own.  As this happens, gas begins to
be trapped between the two bodies; the more effective the
trapping, the more this gas acts to push the bodies apart,
counteracting the asymmetric torque felt by the outer body.
Another difference with respect to Fig. \ref{single_core_omni} is
that the gap-opening body temporarily acquires a substantial
eccentricity, such that its radial excursion becomes equal to the
gap width, where the onset of strong damping then halts
eccentricity growth.  At the same time, the smaller body's
eccentricity decreases temporarily (examination of the 2:1
resonance angles shows a coincident change in their libration
centers). This behavior appears to arise in a fairly narrow range
of parameter space, wherein the torque on the outer body is, for
an extended period of time, just below the threshold needed to
move it from the 1:2 to 2:3 resonance.  It is not apparent in any
of the other simulations presented below, and we defer a closer
examination to future work.

\begin{figure}
\plotone{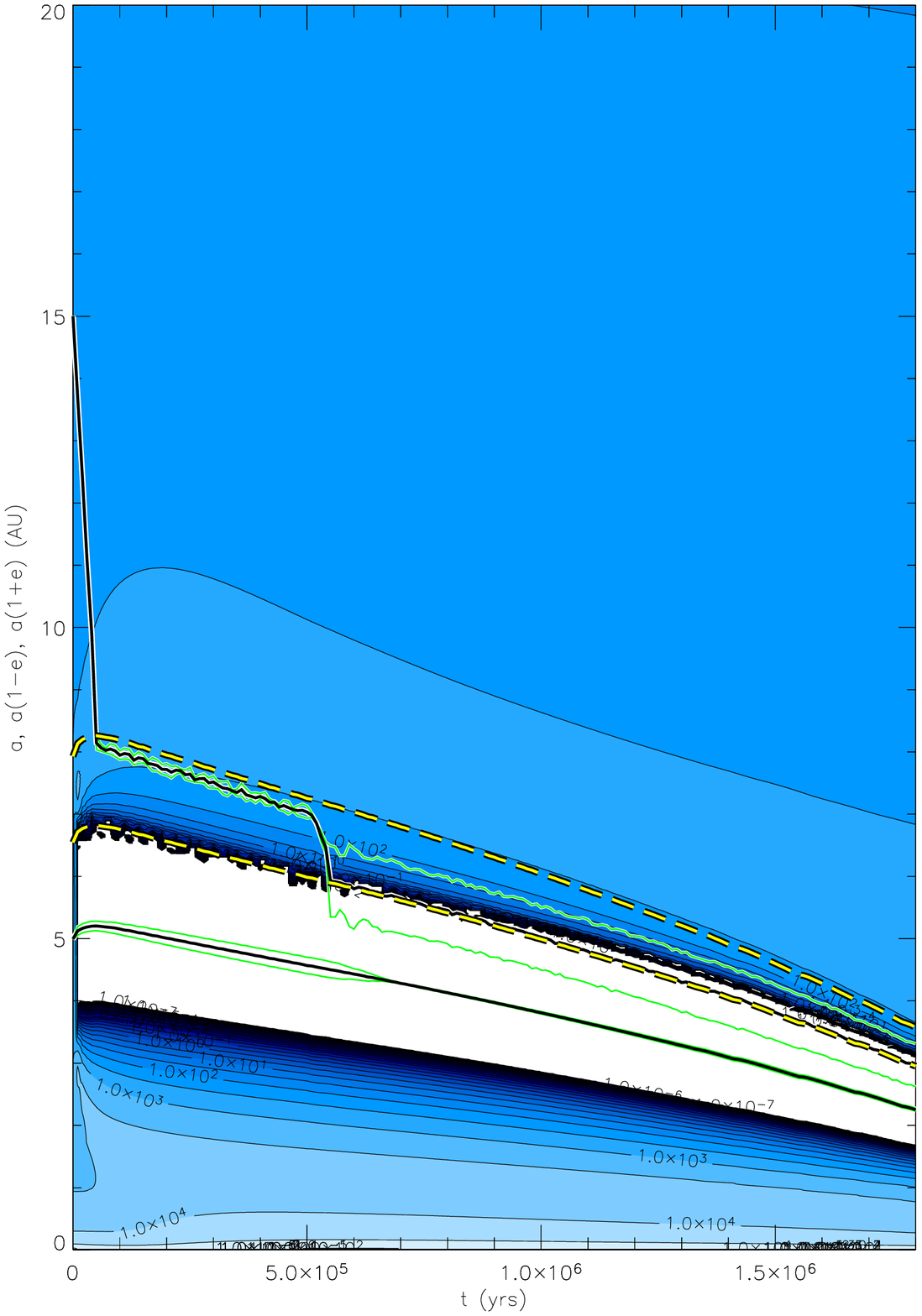}
\caption{The simulation shown in Fig. \ref{idl_paper}, repeated
with a higher initial gas disk surface density of $\Sigma \approx
3000 (r/{\rm 1\,AU})^{-1}\,{\rm g cm^{-2}}$.  This time, the outer
body transitions from the 1:2 to the 2:3 MMR at $5 \times 10^5$
years.} \label{idl_paper_3000}
\end{figure}

We perform a second simulation with a more massive gas disk,
$\Sigma_g \approx 3000 (r/{\rm 1\,AU})^{-1}\,{\rm g cm^{-2}}$. The
results are shown in Fig. \ref{idl_paper_3000}; this time, the
outer planet goes from the 1:2 to the 2:3 resonance with the inner
almost as soon as its growth commences.  This puts it right at the
outer edge of the gap.

\subsection {Multiple protoplanets}

\begin{figure}
\epsscale{0.85} \plotone{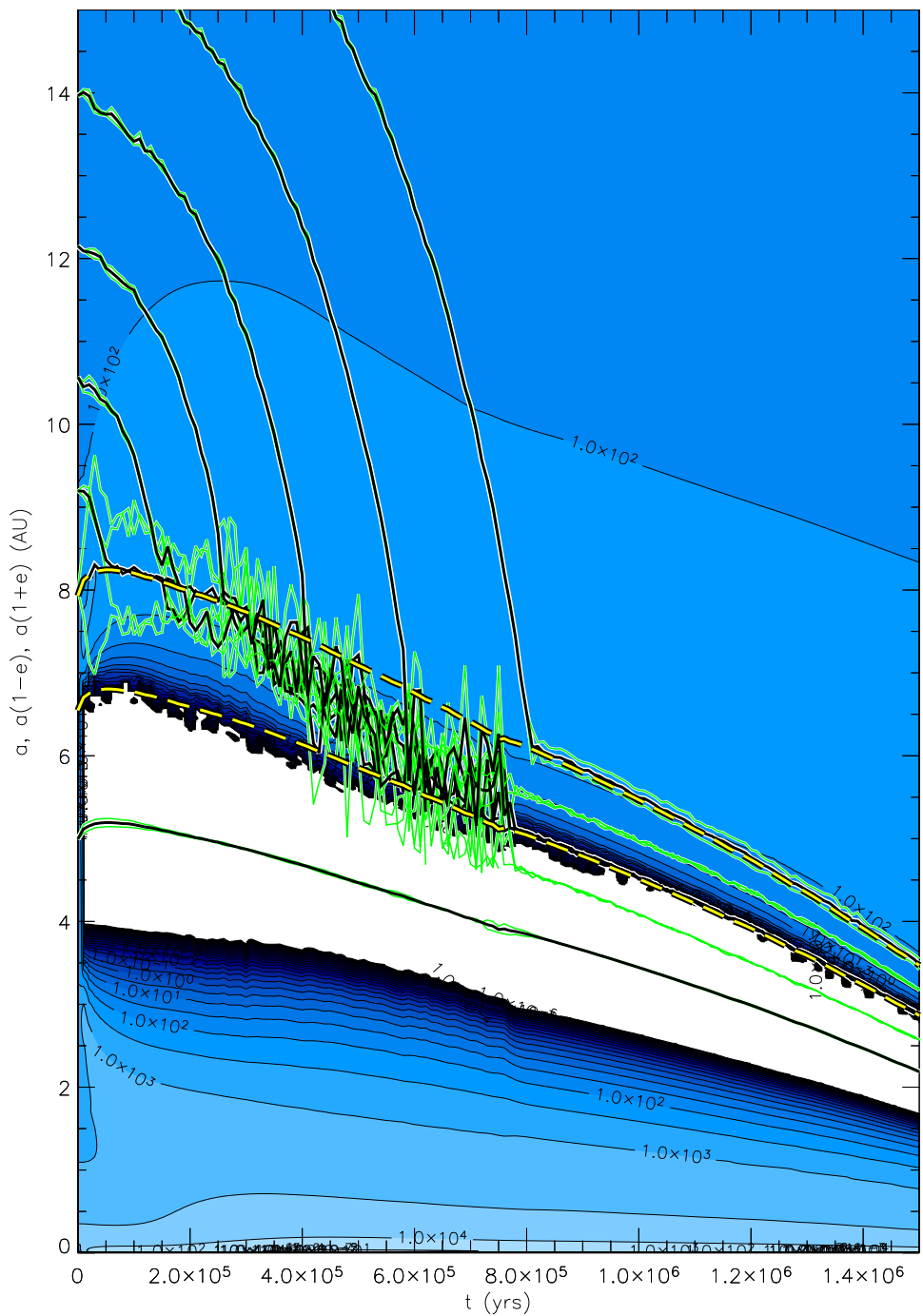}
\caption{Simulation of the orbital evolution of a population of
growing protoplanets outside a gap-opening gas giant, plotted as
in Figs. \ref{idl_paper} and \ref{idl_paper_3000}. The
protoplanets migrate inward rapidly, and one protoplanet after
another encounters the exterior mean-motion resonances of the
giant.  After a period of vigorous interactions lasting until
$\sim 8 \times 10^5$ years, the protoplanets all occupy either the
1:2 or the 2:3 MMR.}\label{idl_paper_multi_5}
\end{figure}

\begin{figure}
\epsscale{1.0} \plotone{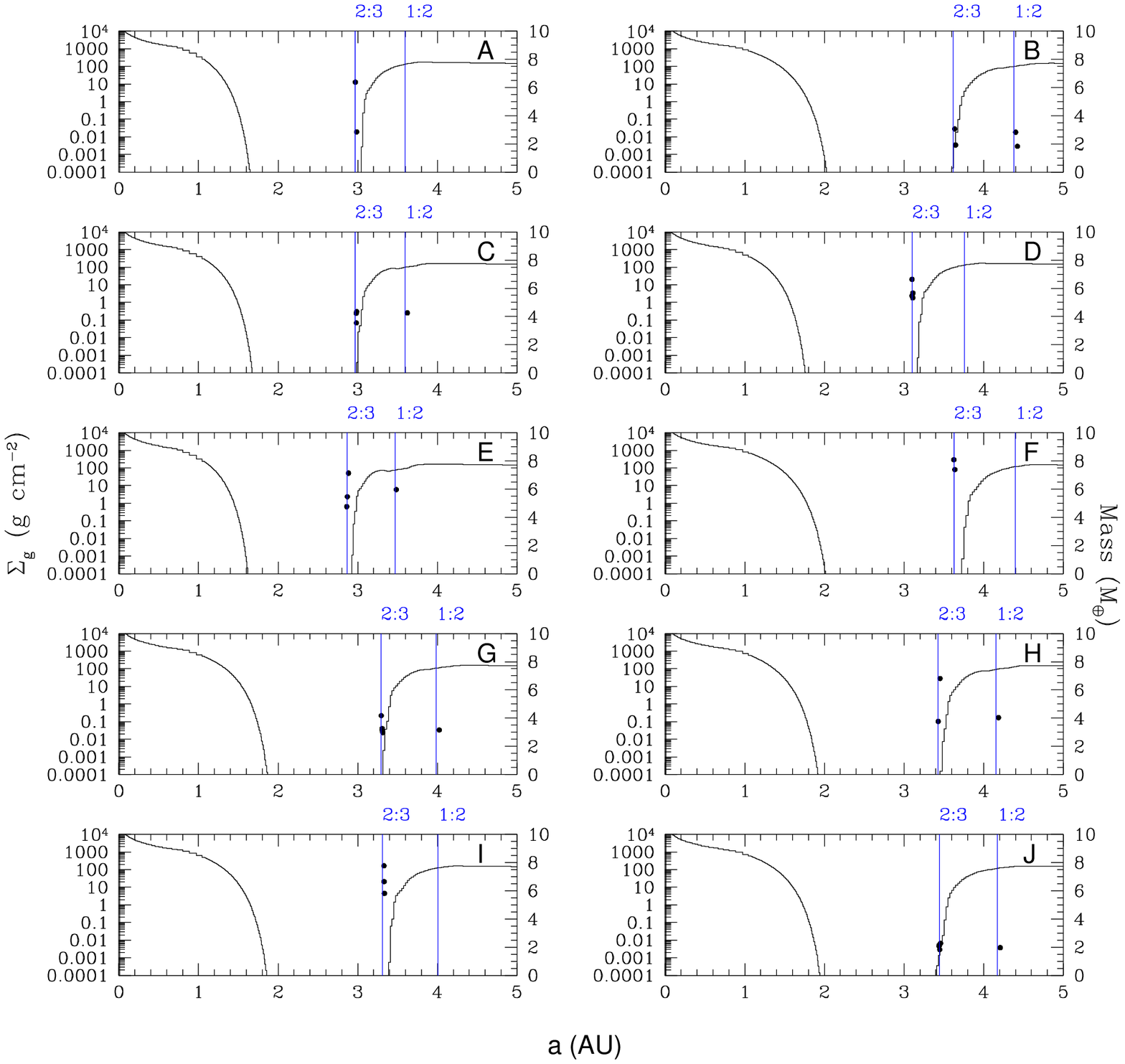}
\caption{Snapshots taken at $1.5 \times 10^6$ years of a set of
ten simulations (differing only in the initial orbital phases of
the bodies) of a population of growing protoplanets on orbits
exterior to a gap-opening gas giant. The locations of the exterior
2:3 and 1:2 mean-motion resonances of the gap-opening planet are
shown as vertical lines.  All surviving protoplanets are locked in
one or the other resonance.  }\label{vdisk_snapshots}
\end{figure}

Using the same prescription as in \S \ref{multiple cores omni}, we
again grow multiple protoplanets originating as $0.1\,M_\oplus$
bodies exterior to the gas giant.  A typical simulation is shown
in Fig. \ref{idl_paper_multi_5}.  In addition, snapshots at $1.5
\times 10^6$ years of a set of similar simulations (differing only
in the randomly-generated initial orbital phases of the bodies)
are shown in Fig \ref{vdisk_snapshots}.  All surviving
protoplanets at this time are locked in either the 2:3 or the 1:2
exterior MMR of the gap-opening giant, with the majority residing
in the former.  As in the simulations of \S \ref{multiple cores
omni}, multiple protoplanets---as many as four in this case---end
up sharing the 2:3 resonance.

\section{Discussion}
\label{discussion}

The numerical simulations we have performed here suggest that when
rapidly-migrating, sub-gap-opening protoplanets approach a
gap-opening gas giant, resonant capture is a likely outcome.  The
two different approaches we use in our simulations---first,
dissipative forces added to an N-body code, and then, a more
self-consistent treatment of the disk which includes gap formation
and viscous accretion---both produce similar results.  A
particularly interesting feature common to both sets of simulation
results is the ``stacking" of multiple multi-Earth-mass bodies in
a single mean-motion resonance of the gas giant.  Such seemingly
precarious resonant configurations can be assembled in our
simulations because the eccentricity damping we apply to sub-gap
bodies stabilizes them against catastrophic mutual scattering with
pre-existing resonant bodies as they approach a MMR.

The ability of a giant planet locked into the viscous transport of
the disk to arrest the orbital decay of O($10^0- 10^1)\,M_\oplus$
 bodies, which, at least in a simple disk model like the one used
here, move inward much faster than the disk material, has
important implications for giant planet formation.  In the classic
core-accretion model
\citep{1996Icar..124...62P}, most of a giant planet's
``childhood", several Myrs in length, is spent as a $\sim
10^1\,M_\oplus$ body waiting to accrete a massive gas envelope.
However, the work of
\citet{1997Icar..126..261W} and
e.g. \citet{2000MNRAS.315..823P},
summarized in \S \ref{background}, implies an orbital decay
timescale for such bodies of only $\sim 10^4$ years. Thus, having
cores held up by a more slowly-migrating gas giant can potentially
be a big help in enabling them to reach core mass at all. However,
even a gap-opening planet does not offer a safe haven
indefinitely.  As can be seen in the simulations of \S
\ref{vdisk}, the gas giant undergoes significant migration in only
2 Myrs, moving in more than 1 AU/Myr on average. And this, it
should be stressed, is with a rather low $\alpha$ viscosity
parameter of $10^{-4}$.  With $\alpha=10^{-3}$, a gap-opening
planet starting at 5 AU has well below 1 Myr before it plunges
into the star, barring some sort of ``parking mechanism" at the
inner edge of the disk (e.g. a central magnetospheric cavity;
\citealp*{1996Natur.380..606L}).  In other words, for a
sufficiently large disk viscosity, Type II migration becomes just
as problematically fast as Type I migration.  Thus, in order for a
gas giant to be of use as a safety net for smaller bodies, it must
exist in a disk, or a region of a disk, that has fairly low
viscosity. Just such a possibility is offered by the layered
accretion or ``dead zone" disk model
\citep{1996ApJ...457..355G,2003ApJ...598..645M}, in which the bulk
of the disk transports angular momentum only at small radii, $\la$
1 AU, and beyond several AU. In between, angular momentum is only
transported in the disk's surface layers. This is because the
assumed source of disk viscosity, magnetorotational instability
\citep{1991ApJ...376..214B}, only operates in a sufficiently
ionized disk; at small radii the star can provide this ionization
and at larger radii cosmic rays take over, but at intermediate
radii, the column depth of the disk is too high to allow cosmic
rays to penetrate all the way through. The result is that from
less than 1 AU to several AU (the exact bounds depend on the
details of the model) the disk has a low viscosity, and thus
gas---and gap-embedded planets---will accrete slowly through this
region, giving planet formation more time to play out.  Even then,
the ultimate survival of whatever planets are produced may require
the removal of at least part of the outer disk by photoevaporation
rather than accretion
\citep*{1993Icar..106...92S,2000prpl.conf..401H,2003ApJ...585L.143M}.
It has been suggested that a giant planet may promote the
accretion of further planets due to the enhancement of disk
surface density near the edges of a freshly-excavated gap
\citep{1979MNRAS.188..191L}. The outer gap edge, in particular,
may locally concentrate solids because the transition from super-
to sub-Keplerian gas orbital speed there ought to causes a pile-up
of planetesimals in a narrow radial range
\citep{2000ApJ...540.1091B}.  We demonstrate that, in agreement
with the suggestion of
\citet{1996LPI....27..479H}, an analogous effect exists when
solids grow to larger size, where disk tides rather than
aerodynamic drag are the primary agent of gas-solids interaction.
Indeed, both effects---the pile-up of small planetesimals and the
resonant capture of large protoplanets---are likely to operate in
concert. This brings up a shortcoming in our prescription for
protoplanet growth:  We use an estimate based on orderly
post-runaway (``oligarchic'') growth, which assumes a smooth
surface density of solids and a uniform spacing of neighbouring
protoplanets.  Both assumptions will certainly not hold up well
when a local enhancement of solids exists, and especially when
multiple bodies are competing to accrete \ in the {\it same}
mean-motion resonance.  However, our intent here has been merely
to make the accretion prescription plausible in the broadest
sense; a more detailed study of the accretion of multiple
resonantly-locked, co-orbital protoplanets is a subject for future
work.

A shortcoming common to both of our simulation schemes is that we
neglect the effect of corotation resonances on sub-gap-opening
planets. Since their strength is proportional to the gradient of
the the quantity ($\Omega \Sigma_g/\kappa^2$), this is more of an
issue for gap-opening planets, which have large $\Sigma_g$
gradients at the gap edges.  These resonances are pivotal in
determining the rate---indeed the sign---of the eccentricity
evolution of planets in gaps
\citep{2003ApJ...585.1024G,2003ApJ...587..398O}. In modeling
gap-opening planets, we also make no explicit use of corotation
resonance torques, but rather use a compromise among the competing
models for the net effect of all resonances on the eccentricity
evolution: Eccentricity is neither driven nor damped while the
planet is fully in its gap, but is damped once it becomes
eccentric enough to run into the gap edges.  A more general caveat
is that our simulation results are contingent on the validity of
treating planet-disk torques of multiple nearby bodies in an
azimuthally-averaged, linearly-superposed manner.  It will be very
interesting to see to what extent full hydrodynamic simulations
reproduce the resonant configurations we have generated here,
especially the extreme case of multiple Type I bodies sharing a
Type II body's MMR.

The simulations performed here lend support to the idea that the
pre-existence of a giant planet makes the formation of subsequent
planets easier.  If so, the formation of the first giant planet
may be the critical event in a protoplanetary disk.  But with no
``safety net" of its own, how can the first solid core be
assembled fast enough?
\citet{2004ApJ...606..520M} offer one possibility:  In a more
complex disk model than is usually used for such calculations
(including the ones in this work), opacity transitions can create
local migration bottlenecks which stretch the Type I timescale to
$\sim 10^6$ years.  The work of
\citet{2004ApJ...614..490C} is also suggestive:  They show that
icy planetesimals undergoing orbital decay due to aerodynamic gas
drag can produce and maintain a massive plume of water vapor once
they reach the water sublimation radius (``snow line''). If much
of this water recondenses just outside the snow line, a large
local enhancement in the surface density of solids may result.  At
that radius, accretion will be more rapid and produce bigger
bodies than elsewhere.  Furthermore, although we have been working
from the perspective of the core accretion model, one can also
imagine a hybrid scenario wherein the first giant planet forms
rapidly
via the competing disk instability model
(\citealp{1978M&P....18....5C}, \citealp{1997Sci...276.1836B},
\citealp{2002Sci...298.1756M}), thus avoiding the Type I issue
altogether, and subsequently facilitates the formation of its
successors by core accretion.
This initial, innermost planet (or planets) may well end up
migrating into the star, leaving only the successors  to populate
the mature system.  That could circumvent some of the
inconsistencies with observations commonly associated with the
disk instability model, namely higher-than-detected planet masses
and (in the context of our Solar System) insufficiently enhanced
planet metallicities.  Finally, regardless of the formation
mechanism, it is worth remembering that only about a quarter of
detected multi-planet systems contain resonant pairs. Therefore,
an important question to be addressed in future work is how
readily mean-motion resonances can be broken as a system evolves.

\section{Summary}
\label{summary} We have performed numerical simulations which
suggest that gap-opening (Type II) gas giant-sized bodies are very
efficient at resonantly capturing sub-gap-opening (Type I) bodies
of Earth to Neptune size in their exterior mean-motion resonances.
In fact, multiple such bodies may accumulate in a single
resonance.  The local concentration of such a large amount of mass
for significant lengths of time stands in sharp contrast to the
problem which arises when considering the core-accretion growth of
a single giant planet in isolation, namely that the growth
timescale exceeds the migration timescale long before the critical
mass for runaway accretion of a gas envelope ($\sim
10^1\,M_\oplus$) is reached.  Hence, the {\it first} gas giant to
form may be the most important, paving the way for its successors.
Also, these results suggest one way to build mean-motion resonant
systems of giant planets like those discovered by radial-velocity
surveys to be orbiting GJ 876, HD 82943, and 55 Cnc: Rather than
giant planets forming separately and then undergoing convergent
migration, resonance locking may happen much earlier, when the
second planet is still just a growing core.

\acknowledgements{I would like to thank G. Bryden, W. Kley, D.
Lin, M.-H. Lee, G. Marcy and F. Masset for stimulating and
informative discussions, the Kavli Institute for Theoretical
Physics for its hospitality during the early part of this work,
and an anonymous referee for valuable comments and suggestions.
This work is supported by the Natural Sciences and Engineering
Research Council of Canada.}

\end{document}